\documentclass[reprint,superscriptaddress,amsmath,amssymb,aps,nofootinbib]{revtex4-2}  
\usepackage{graphicx}
\usepackage{dcolumn}
\usepackage{bm}
\usepackage{color}
\usepackage{epstopdf} 
\usepackage{enumerate}
\usepackage{braket}
\usepackage{siunitx}

\begin{document}
\title{High-field phase diagram of the chiral-lattice antiferromagnet Sr(TiO)Cu$_4$(PO$_4$)$_4$}

\author{Toshihiro Nomura}
\email{tnomura@mail.dendai.ac.jp}
\affiliation{Tokyo Denki University, Adachi, Tokyo 120-8551, Japan}
\affiliation{Hochfeld-Magnetlabor Dresden (HLD-EMFL) and Würzburg-Dresden Cluster of Excellence ct.qmat,  Helmholtz-Zentrum Dresden-Rossendorf, 01328 Dresden, Germany}
\affiliation{Institute for Solid State Physics, The University of Tokyo, Kashiwa, Chiba 277-8581, Japan}
\author{Yasuyuki Kato}
\email{yasuyuki.kato@ap.t.u-tokyo.ac.jp}
\affiliation{Department of Applied Physics, The University of Tokyo, Tokyo 113-8656, Japan}
\author{Yukitoshi Motome}
\affiliation{Department of Applied Physics, The University of Tokyo, Tokyo 113-8656, Japan}
\author{Atsushi Miyake}
\affiliation{Institute for Solid State Physics, The University of Tokyo, Kashiwa, Chiba 277-8581, Japan}
\affiliation{Institute for Materials Research, Tohoku University, Oarai, Ibaraki 311-1313, Japan}
\author{Masashi Tokunaga}
\affiliation{Institute for Solid State Physics, The University of Tokyo, Kashiwa, Chiba 277-8581, Japan}
\author{Yoshimitsu Kohama}
\affiliation{Institute for Solid State Physics, The University of Tokyo, Kashiwa, Chiba 277-8581, Japan}
\author{Sergei Zherlitsyn}
\affiliation{Hochfeld-Magnetlabor Dresden (HLD-EMFL) and Würzburg-Dresden Cluster of Excellence ct.qmat,  Helmholtz-Zentrum Dresden-Rossendorf, 01328 Dresden, Germany}
\author{Joachim Wosnitza}
\affiliation{Hochfeld-Magnetlabor Dresden (HLD-EMFL) and Würzburg-Dresden Cluster of Excellence ct.qmat,  Helmholtz-Zentrum Dresden-Rossendorf, 01328 Dresden, Germany}
\affiliation{Institut f\"ur Festk\"orper- und Materialphysik, TU-Dresden, 01062 Dresden, Germany}
\author{Shojiro Kimura}
\affiliation{Institute for Materials Research, Tohoku University, Katahira 2-1-1, Sendai 980-8577, Japan}
\author{Tsukasa Katsuyoshi}
\affiliation{Department of Advanced Materials Science, The University of Tokyo, Kashiwa, Chiba 277-8561, Japan}
\author{Tsuyoshi Kimura}
\affiliation{Department of Applied Physics, The University of Tokyo, Tokyo 113-8656, Japan}
\affiliation{Department of Advanced Materials Science, The University of Tokyo, Kashiwa, Chiba 277-8561, Japan}
\author{Kenta Kimura}
\email{kentakimura@omu.ac.jp}
\affiliation{Department of Advanced Materials Science, The University of Tokyo, Kashiwa, Chiba 277-8561, Japan}
\affiliation{Department of Materials Science, Osaka Metropolitan University, Osaka 599-8531, Japan}

\date{\today}

\begin{abstract}
We studied the high-field phase diagram of a chiral-lattice antiferromagnet Sr(TiO)Cu$_4$(PO$_4$)$_4$ by means of ultrasound, dielectric, and magnetocaloric-effect measurements.
These experimental techniques reveal two new phase transitions at high fields, which have not been resolved by previous magnetization experiments.
Specifically, the $c_{66}$ acoustic mode shows drastic changes with hysteresis for magnetic fields applied along the $c$ axis, indicating a strong magneto-elastic coupling.
Combined with cluster mean-field theory, we discuss the origin of these phase transitions.
By considering the chiral-twist effect of Cu$_4$O$_{12}$ cupola units, which is inherent to the chiral crystal structure, the phase diagram is reasonably reproduced.
The agreement between experiment and theory suggests that this material is a unique quasi-two-dimensional spin system with competing exchange interactions and chirality, leading to a rich phase diagram.
\end{abstract}

\maketitle 

\section{Introduction}

\begin{figure*}[tb]
\centering
\includegraphics[width=0.99\linewidth]{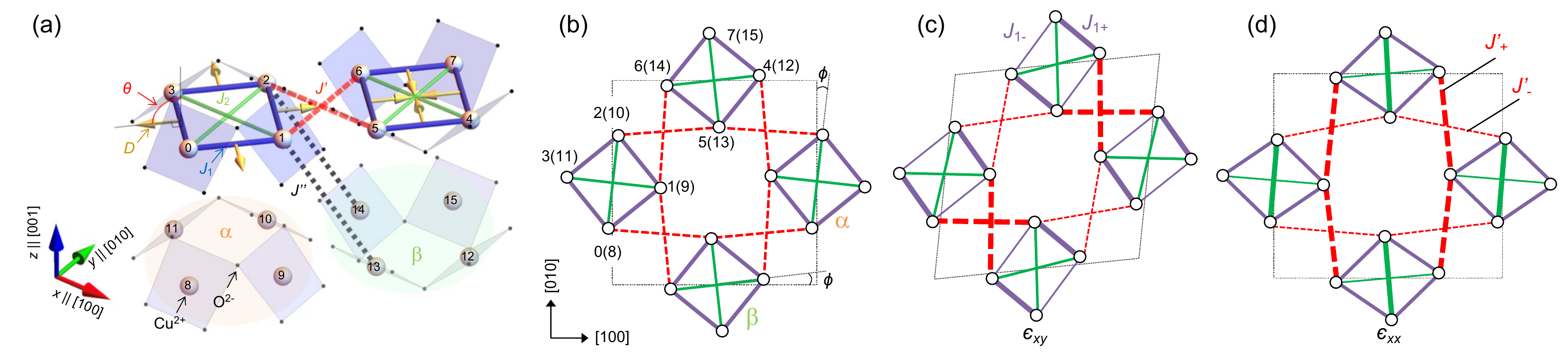}
\caption{
(a) Crystal structure of SrTCPO with magnetic interactions; nearest-neighbor exchange ($J_1$), next-nearest-neighbor exchange ($J_2$), inter-cupola exchanges ($J'$ and $J''$), and DM vector ($D$). $\theta$ is the angle of the DM vector from the $c$ axis. The site index $\ell$ (0--15) is denoted on each Cu$^{2+}$.
(b) In-plane interactions of SrTCPO with the twist angle $\phi$.
(c, d) Strained unit cells with (c) $\epsilon_{xy}$ and (d) $\epsilon_{xx}$.
The shorter bonds are shown by thicker lines.
} 
\label{fig1}
\end{figure*}

Chiral magnets are of great interest because of their exotic properties such as helical magnetism, multiferroicity, and skyrmionic textures \cite{MnSi76,MnSi09,Seki12,Kanazawa16}.
Dzyaloshinskii-Moriya (DM) interaction, inherent to the chiral structure without inversion symmetry, leads to  canted spin textures, which are rarely observed in usual magnets.
Helical magnetism can also be realized in the presence of geometrical frustration (or competing exchange interactions), hindering trivial long-range order \cite{Yamasaki13,Tsurkan13}.
In the presence of quantum fluctuations, exotic ground states such as quantum spin liquids and magnon Bose-Einstein condensates may emerge \cite{Balents,Giamarchi}.
A system with chirality and geometrical frustration is an attractive platform for novel exotic states of matter, although the appropriate material design is challenging.

A series of chiral-lattice crystals, $A$(TiO)Cu$_4$(PO$_4$)$_4$ ($A$TCPO with $A$ = Ba, Sr, Pb) with space group $P42_12$ show fascinating properties originating from the Cu$_4$O$_{12}$ square-cupola units \cite{KimuraIC16,KimuraPRM18} [Fig.~1(a)].
The upward ($\alpha$) and downward ($\beta$) square cupolas alternatively align in the $ab$ plane, and the compounds can be regarded as quasi-two-dimensional systems.
On each Cu$_4$O$_{12}$ unit, the four $S=1/2$ spins of the Cu$^{2+}$ ions form a sort of spin tetramer and can host magnetic multipole moments (monopole, quadrupole, and toroidal moments) \cite{KimuraNC16,Kato17,Kato19}. 
Under external magnetic fields, these magnetic multipoles order and/or disorder in a ferroic or antifferoic manner, leading to successive magnetic phase transitions.
Since these magnetic multipoles simultaneously break space-inversion and time-reversal symmetries,  magnetoelectric (ME) coupling appears in $A$TCPO.
Recent experimental and theoretical studies revealed a relation between the magnetic structure and the ME responses in $A$TCPO 
\cite{KimuraNC16,Babkevich17,Kato17,KimuraPRM18,KimuraPRB18,Kato19,Islam18,Kimura19,Rasta20,Katsuyoshi21}.
Noteworthy, the cluster mean-field (CMF) approach has clarified the evolution of the magnetic multipoles in the Cu$_4$O$_{12}$ units as a function of magnetic field and allowed to predict the ME responses.
However, the experimental study on the magnetic field--temperature ($B$--$T$) phase diagram of $A$TCPO up to the saturation field has not yet been completed.

In this study, we present the experimental $B$--$T$ phase diagram of SrTCPO using the combined results of  ultrasound, dielectric, and magnetocaloric-effect (MCE) measurements.
SrTCPO orders antiferromagnetically at $T_\mathrm{N}=6.3$~K, and an antiferroic ME response is simultaneously observed in the dielectric susceptibility \cite{KimuraNC16}.
The high-field magnetization results suggest that the magnetic saturation field $B_\mathrm{s}$ for the  crystallographic [001], [100], and [110] directions are $B_\mathrm{s}^{[001]}=37.0$~T, $B_\mathrm{s}^{[100]}=41.0$~T, and $B_\mathrm{s}^{[110]}=38.9$~T, respectively \cite{Kato17}.
Before the saturation fields, spin-flop-like transitions are observed at $B_\mathrm{c1}^{[001]}=27.4$~T, $B_\mathrm{c1}^{[100]}=13.5$~T, and $B_\mathrm{c1}^{[110]}=15.0$~T, respectively \cite{Kato17}.
Our experimental techniques allow us to detect two additional phase transitions which are not observed in the magnetization.
We discuss the origin of these new phase transitions based on the CMF theory.

This paper is organized as follows.
In Sec. II, we explain our experimental and theoretical methods.
In Sec. III, we present the experimental results of the ultrasound, dielectric, and MCE measurements together with the experimental $B$--$T$ phase diagram.
In Sec. IV, we show the results of theoretical calculations corresponding to the experiments.
In Sec. V, we discuss the phase diagram by comparing the experimental and theoretical results.
In Sec. VI, conclusive remarks are given.

\section{Methods}
\subsection{Experimental methods}
We grew single crystals of SrTCPO by the flux method \cite{KimuraIC16}.
We performed powder x-ray diffraction (XRD) on crushed single crystals and confirmed a single phase. 
We determined the crystal orientation by Laue x-ray scattering.
The crystal used in this study has L-type chirality, determined by optical-activity measurement at a wavelength of 450 nm.
The size of the investigated crystal was $4\times4\times2$ mm$^3$.

We performed ultrasound-velocity measurements based on the transmission pulse-echo technique with a phase-sensitive detection \cite{Zherlitsyn14,Luthi,Kohama22}.
We attached two LiNbO$_3$ transducers (41$^\circ$-X cut for transverse and 36$^\circ$-Y cut for longitudinal acoustic modes) to the polished surfaces of the single crystal.
We analyzed the phase of the echo signal and obtained the relative change in the sound velocity.
The ultrasound frequency was typically 20--180 MHz.
For some experimental geometries, the ultrasound frequency was increased up to 500 MHz to investigate the nonreciprocal properties (magnetochiral effect) \cite{Nomura19,Nomura23,Rikken97}.
The elastic modulus $c_{ij}$ was calculated as $c_{ij}=\rho v_{ij}^2$, using the density $\rho=4.109$ g/cm$^3$ \cite{KimuraIC16}.
The density change below 200 K is negligibly small \cite{Islam18}. %around 0.3 \%.
The uncertainty of $c_{ij}$ is around $\pm$5 \%.
The experimental geometries (propagation and polarization vectors, ${\bf k}$ and ${\bf u}$, respectively) for each acoustic mode are summarized in Table 1 with the sound velocity and the calculated elastic modulus at 2 K.
Here, $c_\mathrm{T}=(c_{11}-c_{12})/2$.

\begin{table}
\caption{\label{tab:table1} Experimental geometries for each acoustic mode. ${\bf k}$ and ${\bf u}$ are the propagation and polarization vectors, respectively. The sound velocity $v_{ij}$ and elastic modulus $c_{ij}$ at 2 K are summarized. The irreducible representations (IR) for $D_4$ symmetry and the related strains are also shown.}
\begin{ruledtabular}
\begin{tabular}{lccccc|cc}
& ${\bf k}$ & ${\bf u}$ & $v_{ij}$ & $c_{ij}$ && IR & Strain \\
& & & (km/s) & (GPa) && &  \\
\hline
$c_{66}$&[100]&[010]& 3.32 & 45 && $B_2$ & $\epsilon_{xy}$ \\
$c_{11}$&[100]&[100]& 6.47 & 172 && $A_1 \oplus B_1$ & $\epsilon_{xx}$, $\epsilon_{yy}$ \\
$c_\mathrm{T}$&[110]&[1$\overline{1}$0]& 4.20 & 78 && $B_1$ & $\epsilon_{xx}-\epsilon_{yy}$ \\
$c_{44}$&[100]&[001]& 2.41 & 24 && $E$ & $\epsilon_{yz}$, $\epsilon_{zx}$ \\
$c_{33}$&[001]&[001]& 4.47 & 82 && $A_1$ & $\epsilon_{zz}$ \\
\end{tabular}
\end{ruledtabular}
\end{table}

We measured the dielectric constant $\varepsilon$ up to 25 T by using a superconducting magnet and up to 50 T in a pulsed magnet.
For these measurements we utilized an LCR meter (Agilent E4980) and a capacitance bridge for the static- and pulsed-field experiments, respectively \cite{Miyake20}.
The pulse duration for the dielectric experiment was $\sim 35$~ms, which was shorter than that of the ultrasound experiment $\sim 150$~ms.
Because of the shorter duration, the temperature of the crystal changed during the pulsed magnetic fields.
Therefore, we used only the static-field data for the phase diagram.

We obtained the adiabatic temperature change in the pulsed magnetic field (magnetocaloric effect, MCE) using a RuO$_2$ thermometer (900 $\Omega$, $0.6\times0.3\times0.1\ \mathrm{mm}^3$) \cite{13RSI_Kihara,17PRB_Nomura,Yamamoto21}.
We glued the thermometer on the polished surface of the single crystal ($4\times4\times2\ \mathrm{mm}^3$).
The sample was placed in vacuum to ensure adiabatic conditions.
We measured the resistance of the thermometer using a standard ac four-probe method and a numerical lock-in technique at a frequency of 50 kHz, typical for pulsed-field experiments.

\subsection{Theoretical methods}\label{sec:theoretical_procedures}
We consider an effective spin model for the $S=1/2$ spin degrees of freedom of the Cu$^{2+}$ cations,
which was first introduced for BaTCPO~\cite{Kato17} and later applied to PbTCPO~\cite{KimuraPRM18} and SrTCPO~\cite{Kato19}.
The Hamiltonian reads
\begin{eqnarray}
\mathcal{H}
=\sum_{\langle i,j \rangle} \left[
J_1 {\bf S}_i \cdot {\bf S}_j
- {\bf D}_{ij}
\cdot 
({\bf S}_i \times {\bf S}_j)
\right]
+ J_2 \sum_{\langle\langle i,j \rangle\rangle} {\bf S}_i \cdot {\bf S}_j \nonumber\\
+ J' \sum_{(i,j)} {\bf S}_i \cdot {\bf S}_j
+ J'' \sum_{((i,j))} {\bf S}_i \cdot {\bf S}_j
-g\mu_{\rm B}\sum_i {\bf B}\cdot{\bf S}_i, ~~
\label{eq:model}
\end{eqnarray}
where ${\bf S}_i=(S^x_i, S^y_i , S^z_i)$ represents the $S=1/2$ spin at site $i$,
$J_1$, $J_2$, $J'$, and $J''$ represent four dominant antiferromagnetic exchange interactions~[Fig.~\ref{fig1}(a)],
${\bf D}_{ij}$ is the Dzyaloshinskii-Moriya vector,
and the last term represents the Zeeman coupling with an isotropic $g$-factor $g$ and the Bohr magneton $\mu_{\rm B}$.
The sums for $\langle i,j \rangle$, $\langle\langle i,j \rangle\rangle$,
$(i,j)$, and $((i,j))$ run over the $J_1$, $J_2$, $J'$, and $J''$ bonds, respectively.
We use the model parameters for SrTCPO found in Ref.~\cite{Kato19} as $J_1=0.6$, $J_2=1/6$, $J'=1/2$, $J''=1/100$, $D=0.7$, and $\theta=90^\circ$.
While the effect of the chiral twist of the square cupolas has been neglected in Ref.~\cite{Kato19},
we take into account a nonzero chiral twist to explain the magnetic transitions found in this study in the high-field regime for ${\bf B}\parallel [100]$.
In the present calculations, we take $\phi = 1^{\circ}$ \cite{note1},
which leads to a $\pm 1^\circ$ tilt of ${\bf D}_{ij}$ from the $\langle 110 \rangle$ directions~[Figs.~\ref{fig1}(a) and  \ref{fig1}(b)].

As in the previous studies~\cite{Kato17,KimuraPRM18,Kato19},
we performed calculations based on the cluster mean-field (CMF) theory, which is suitable for cluster-based magnetic materials.
In the CMF method,
the intracupola interactions and the Zeeman coupling are dealt with by exact diagonalization,
and, therefore, quantum fluctuations in each cupola are fully taken into account; the other weaker intercupola interactions ($J'$ and $J''$ terms) are taken into account
by means of conventional mean-field approximation.
Following the previous studies~\cite{Kato17,Kato19}, we consider four square cupolas allocated as shown in Fig.~\ref{fig1}(a) in the CMF method, namely we consider 16 sublattices.

To characterize the magnetically ordered phases, we consider the same order parameter as defined in the previous studies~\cite{Kato17,KimuraPRM18,Kato19}:
\begin{align}
{\bf m}_{\rm AF} = \frac{1}{N_{\rm spin}} \sum_{\ell} (-1)^\ell p_\ell \langle {\bf S}_\ell \rangle,
\end{align}
where $\langle {\bf S}_\ell \rangle $ is the expectation value of the spin operator ${\bf S}_\ell$,
$p_\ell = +1 (-1)$ for the upper (lower) layer in Fig.~\ref{fig1}(a)
and $N_{\rm spin}$ is the number of spins.
We also compute the entropy $S$ that is obtained from the temperature integral of the specific heat as
\begin{align}
S(T) = \int_{T_{0}}^T \frac{C(\tilde{T})}{\tilde{T}} d\tilde{T},
\end{align}
with a sufficiently low $T_0 = 0.01$,
where $C(T)$ is estimated by a derivative of the cubic spline interpolation of the internal energy as $C(T)= (1/N_{\rm spin}) d\langle \mathcal{H} \rangle/dT$.

The elastic constants $c_{66}$ and $c_{11}$ are given by the second derivative of the free energy in terms of the strain as $c=\partial^2 F/\partial \epsilon^2$.
Since the spin interactions linearly change with the strain in general,
we consider the second derivatives of the free energy with respect to the interactions instead.
We take into account a deformation of $J_{1\pm}$ bonds [Fig.~\ref{fig1}(c)] as $J_1 \to (1\pm r)J_1 $, ${\bf D}_{ij} \to (1\mp r){\bf D}_{ij}$ for the strain of $\epsilon_{xy}$
and a deformation of $J'_{-}$ bonds [Fig.~\ref{fig1}(d)] as $J'_{-} \to J' - j' $ for the strain of $\epsilon_{xx}$.
For simplicity, it is assumed that $J_1$ and ${\bf D}_{ij}$ have the same rate of change for $\epsilon_{xy}$,
and the effects on the other interactions by the strain are also neglected.
Then, we compute the second derivatives of the free energy as
\begin{align}
\Delta_r^{(2)} =& \frac{1}{N_{\rm spin}}\frac{\partial^2 F}{\partial r^2},\\
\Delta_{j'}^{(2)} =& \frac{1}{N_{\rm spin}} \frac{\partial^2 F}{\partial {j'}^2}.
\end{align}
One might think that the deformation of $J'$ bonds for the strain of $\epsilon_{xy}$ is also important as indicated in Fig.~\ref{fig1}(c).
However, the effect is canceled out in the CMF calculation and, therefore, is expected to be not significant.
In practical calculations, the first derivatives,
\begin{align}
\Delta_r^{(1)} =&
\frac{1}{N_{\rm spin}}\frac{\partial F}{\partial r},\\
\Delta_{j'}^{(1)} =&
\frac{1}{N_{\rm spin}} \frac{\partial F}{\partial j'},
\end{align}
are obtained by linear combinations of $\langle {\bf S}_i \cdot {\bf S}_j \rangle$ or $\langle {\bf S}_i \rangle \cdot \langle {\bf S}_j \rangle$, and do not require numerical derivatives;
the second derivatives are computed by numerical derivatives of $ \Delta_{r}^{(1)}$ and $ \Delta_{j'}^{(1)}$ as
\begin{align}
\Delta_{r}^{(2)} =&
\frac{1}{2\delta}\left[
\left. \Delta_{r}^{(1)}\right|_{r=\delta} - \left.\Delta_{r}^{(1)} \right|_{r=-\delta}
\right],\\
\Delta_{j'}^{(2)} =&
\frac{1}{2\delta}\left[
\left. \Delta_{j'}^{(1)}\right|_{j'=\delta} - \left.\Delta_{j'}^{(1)} \right|_{j'=-\delta}
\right],
\end{align}
with a sufficiently small $\delta = 10^{-4}$.

\section{Experimental results}
\subsection{Temperature dependence of the acoustic properties}

\begin{figure}[tb]
\centering
\includegraphics[width=0.99\linewidth]{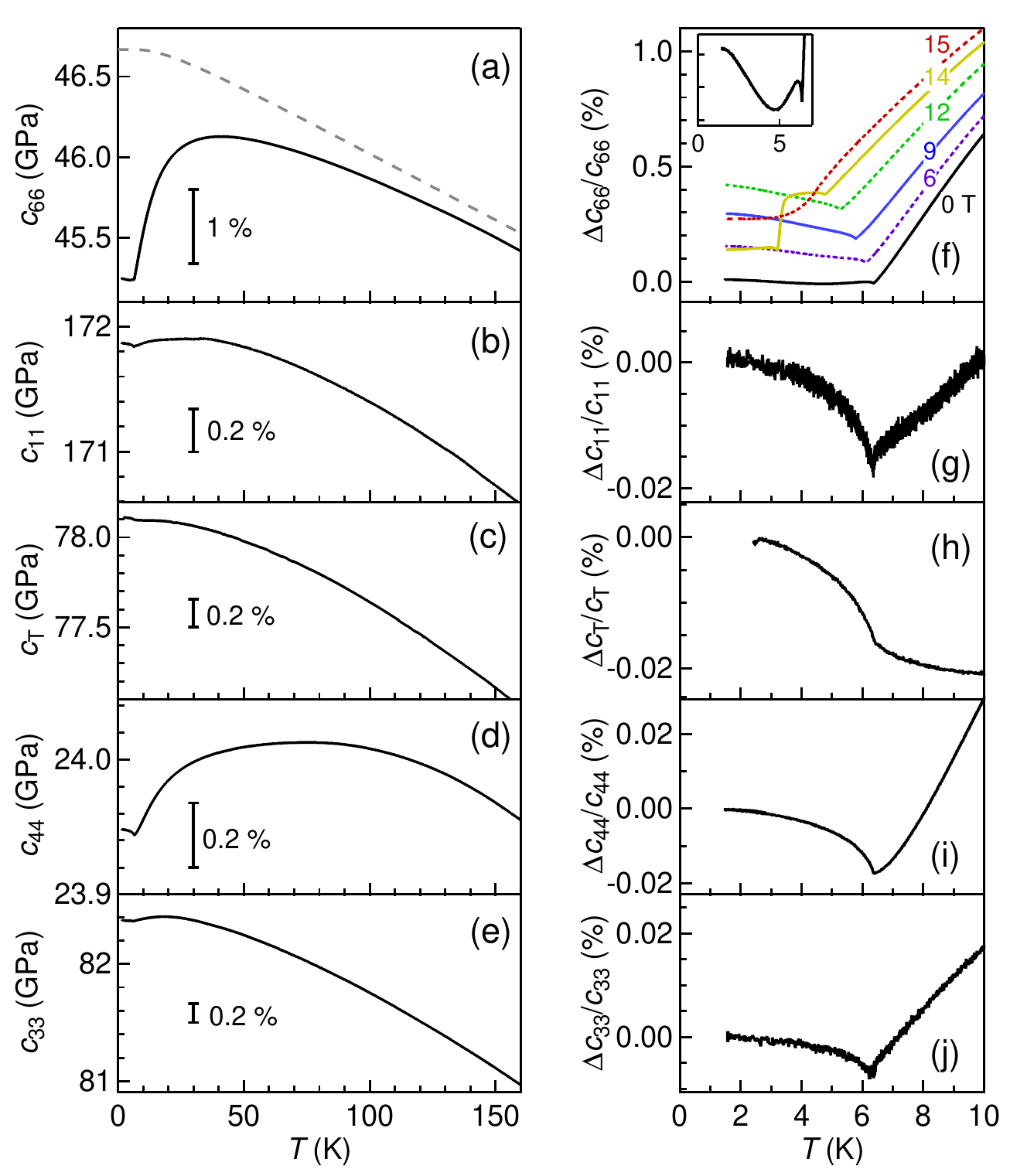}
\caption{
(a--e) Temperature dependence of the elastic constants.
Solid lines show the results at zero field.
The gray dashed line shows the fitted background due to the anharmonic phonon contribution based on Ref. \cite{Varshni70}.
(f--j) Relative change of the elastic constants $\Delta c/c$ near $T_\mathrm{N}$.
(f) Results for zero and applied fields ($\bf{B} \parallel \bf{k} \parallel [100]$) as labeled for each curve. 
The inset shows the enlarged result at zero field.
} 
\label{US_static}
\end{figure}

Figures \ref{US_static}(a)--\ref{US_static}(e) show the temperature dependence of each elastic constant at zero field.
All acoustic modes exhibit an anomaly at $T_\mathrm{N}=6.3$~K, evidencing the long-range magnetic ordering.
We comment that the acoustic attenuation does not show any anomaly at $T_\mathrm{N}$ for all acoustic modes within our experimental resolution. 
The elastic softening towards $T_\mathrm{N}$ reflects developing short-range correlations via magneto-elastic coupling.
The $c_{66}$ mode shows the largest softening of 1.9 \%, indicating the strong spin-lattice coupling with the $ab$-plane shear strain $\epsilon_{xy}$.
The softening of the other acoustic modes are one or two orders of magnitude smaller than that of the $c_{66}$ mode.
The background curve of the $c_{66}$ mode due to the phonon anharmonicity is estimated by the dashed line in Fig.~\ref{US_static}(a) \cite{Luthi,Varshni70}.

Generally, for two-dimensional spin systems, the in-plane acoustic modes ($c_{66}$, $c_{11}$, $c_\mathrm{T}$) show larger spin-lattice coupling than the acoustic modes with out-of-plane strains ($c_{44}$, $c_{33}$) \cite{Luthi,Wolf01,Nomura22}.
This is because the in-plane strains ($\epsilon_{xy}$, $\epsilon_{xx}$, $\epsilon_{xx}-\epsilon_{yy}$) may linearly modulate the Cu-Cu distance, while the out-of-plane strains ($\epsilon_{yz}$, $\epsilon_{zz}$) only modulate with higher-order coupling.
Thus, the weaker elastic anomalies for the $c_{44}$ and $c_{33}$ modes are understandable.
The differences between the in-plane modes ($c_{66}$, $c_{11}$, $c_\mathrm{T}$) reflect how the corresponding strains are coupled to the spin degrees of freedom.

Next, we focus on the anomalies of the elastic constants near $T_\mathrm{N}$ in Figs. \ref{US_static}(f)--\ref{US_static}(j).
For the $c_{11}$, $c_\mathrm{T}$, $c_{44}$, and $c_{33}$ modes, the elastic constants show a single anomaly at the ordering temperature.
In contrast, the $c_{66}$ mode at zero field [enlarged in the inset of Fig.~\ref{US_static}(f)] shows a sharp minimum at $T_\mathrm{N}$ and a broad minimum at around 5 K.
The hardening of $c_{66}$ below $T_\mathrm{N}$ is smaller than for the other modes.
When the magnetic field is applied with $\textbf{B}\parallel\textbf{k} \parallel [100]$, the broad minimum gradually disappears towards 12~T.
At higher magnetic fields, $c_{66}(T)$ shows qualitatively different behavior due to a field-induced phase transition, which is discussed later together with the phase diagram.

\begin{figure}[tb]
\centering
\includegraphics[width=0.9\linewidth]{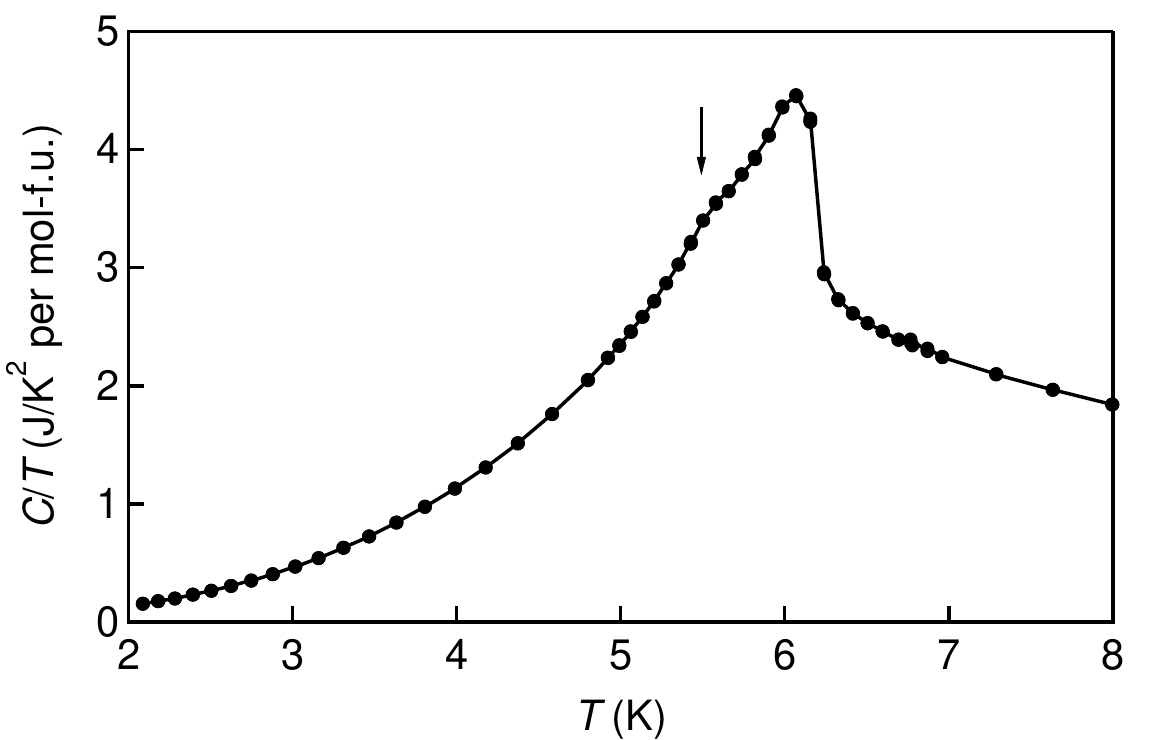}
\caption{
Temperature dependence of the specific heat divided by temperature $C/T$ at zero field.
A broad hump is indicated by the arrow.
} 
\label{c_t}
\end{figure}

Here, we discuss the origin of the broad minimum observed for $c_{66}$ in zero field [Fig.~\ref{US_static}(f)].
Below $T_\mathrm{N}$ in zero field, no phase transition was reported from magnetic-susceptibility, heat-capacity, and dielectric measurements \cite{Islam18,Kato19}. 
However, the spin-lattice relaxation rate of $^{31}$P-NMR exhibits a small discontinuity at around 5~K \cite{Islam18}.
Figure \ref{c_t} shows the $T$ dependence of the specific heat divided by $T$ at zero field, which is obtained by combining the data taken from  Refs. \cite{KimuraPRB18,Kimura20} and new data taken with a narrower $T$ interval around $T_\mathrm{N}$. 
There is a clear hump around 5.5 K.
As discussed later in connection with the MCE results, irreversible heating is observed only below 5~K.
Such irreversible heating indicates slow dynamics of this system.

A similar temperature dependence of the elastic constants below $T_\mathrm{N}$ is also observed for the frustrated magnet GeCo$_2$O$_4$ \cite{Watanabe08,Watanabe11} and the charge-ordering system $\alpha'$-NaV$_2$O$_5$ \cite{Schwenk99}.
In these materials, the characteristic hardening below $T_\mathrm{N}$ was attributed to domain-wall stress and to spin-singlet to triplet excitations.

Here, we discuss the effect of possible magnetic domains in SrTCPO.
The magnetic order at zero field is of antiferro-quadrupolar type ($q_{x^2-y^2}$), where $q_{x^2-y^2}$ magnetic quadrupoles align in a ferroic manner in the two-dimensional magnetic layer, but in an antiferroic manner along the $c$ axis. 
For simplicity, we consider only the single magnetic layer in the $ab$ plane.
The upward ($\alpha$) and downward ($\beta$) cupola alternatively take ``udud" and ``dudu" spin configurations, where ``u" and ``d" denote up and down spin configurations, respectively.
Note that the opposite configuration (``udud" for $\beta$ and ``dudu" for $\alpha$) has the same energy.
Thus, two domains with opposite magnetic quadrupoles can coexist and form magnetic domains.
At low temperatures, the domain motions are restricted due to pinning and possibly affect the elastic properties.
The anomaly at around 5~K could be related to the freezing of the antiferro-quadrupolar domains.

\begin{figure*}[ht]
\centering
\includegraphics[width=0.98\linewidth]{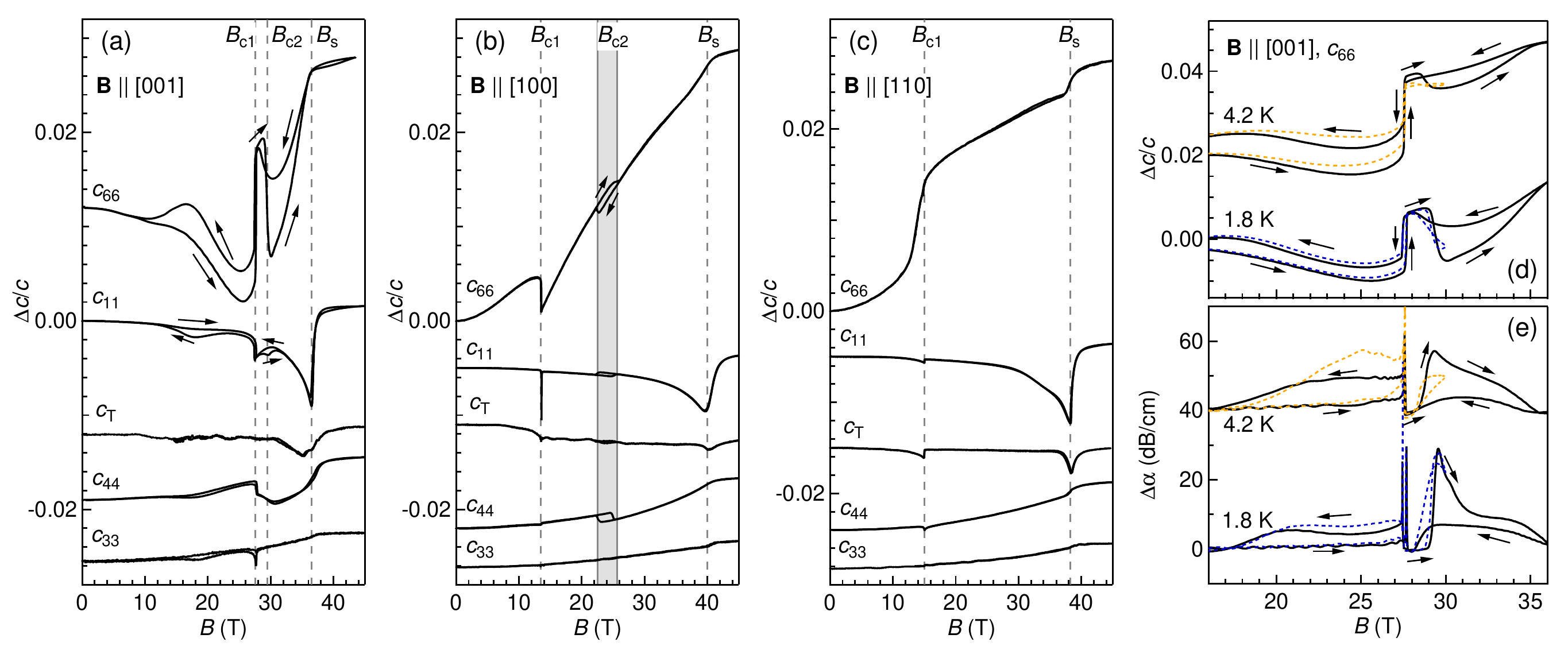}
\caption{
(a-c) Relative changes of the elastic constants $\Delta c/c$ at 1.8 K as a function of magnetic field for the field directions (a) ${\bf B} \parallel [001]$, (b) ${\bf B} \parallel [100]$, and (c) ${\bf B} \parallel [110]$.
Enlarged results of (d) $\Delta c/c$ and (e) the relative change of the acoustic attenuation of the $c_{66}$ mode ($f=122$ MHz) with ${\bf B} \parallel [001]$.
The dashed lines show results obtained using lower maximum-field pulses.
The results are shifted for clarity.
} 
\label{US_pls}
\end{figure*}

\subsection{Magnetic-field dependence of the acoustic properties}

Figure \ref{US_pls} summarizes the relative changes of the elastic constants $\Delta c/c$ at 1.8~K up to the saturation magnetic fields along the [001], [100], and [110] axes.
For each direction, one or two field-induced phase transitions are observed as denoted by the gray dashed lines and the gray band ($B_{\mathrm{c2}}^{[100]}$).
We also performed  measurements at 4.2~K in liquid $^4$He, however, the results show extrinsic hysteresis due to a strong MCE (not shown).
Only below 2~K, where the sample was immersed in superfluid $^4$He, an isothermal condition was realized. 
The hysteresis observed in Fig.~\ref{US_pls} is considered to be intrinsic.

First, the results for ${\bf B} \parallel [001]$ [Fig.~\ref{US_pls}(a)] are discussed.
We observed clear discontinuities of $\Delta c/c$ at $B_{\mathrm{c}1}^{[001]}=27.6$~T, $B_{\mathrm{c}2}^{[001]}=29.5$~T, and $B_{\mathrm{s}}^{[001]}=36.5$~T.
In magnetization measurements, the field-induced transition at $B_{\mathrm{c}1}^{[001]}$ and saturation of the magnetization at $B_{\mathrm{s}}^{[001]}$ were observed \cite{Kato19}.
In $c_{66}$, $c_{11}$, and $c_{44}$, we clearly detect another transition at $B_{\mathrm{c}2}^{[001]}$ with  remarkable hysteresis.
In particular, the $c_{66}$ mode ($\epsilon_{xy}$) shows drastic changes with a complex hysteresis loop, indicating strong in-plane spin-strain correlations.
Figures \ref{US_pls}(d) and \ref{US_pls}(e) show the field dependence of $\Delta c_{66}/c_{66}$ and the relative change of the acoustic attenuation $\Delta \alpha$ near $B_{\mathrm{c}2}^{[001]}$.
Here, results for different peak fields $B_\mathrm{max}$ are presented for 4.2 and 1.8~K.
The complex hysteresis loops are reproduced for these conditions and observed in $\Delta \alpha$ as well.
Note, that the acoustic attenuation becomes minimal at the fields between $B_{\mathrm{c}1}^{[001]}$ and $B_{\mathrm{c}2}^{[001]}$, indicating that the state in this region is a thermodynamically stable phase, not just a transient state.

Second, we discuss the results for ${\bf B} \parallel [100]$ [Fig.~\ref{US_pls}(b)].
A field-induced transition at $B_{\mathrm{c}1}^{[100]}=13.5$ T and saturation of the magnetization at $B_{\mathrm{s}}^{[100]}=40.0$ T are clearly observed.
In addition to these anomalies, a slight change of elasticity is observed at $B_{\mathrm{c}2}^{[100]}\sim 24$~T with hysteresis as indicated by the gray band.
The magnetization shows only a tiny anomaly at $B_{\mathrm{c}2}^{[100]}$ \cite{Kato19}.
At $B_{\mathrm{c}2}$, none of the acoustic modes shows a drastic change, indicating that the symmetry of the magnetic state only slightly changes at this critical field.
Therefore, this phase transition is related to a slight rearrangement of spins in the cupola.

Third, we discuss the results for ${\bf B} \parallel [110]$ [Fig.~\ref{US_pls}(c)].
Again, a field-induced transition at $B_{\mathrm{c}1}^{[110]}=15.0$ T and saturation of the magnetization at $B_{\mathrm{s}}^{[110]}=38.2$ T are observed.
For this field direction, no additional feature is detected compared to the magnetization results \cite{Kato19}.

Last, we note on the possible nonreciprocal acoustic properties in this system \cite{Nomura19,Nomura23,Rikken97}.
For the Faraday geometries (${\bf B} \parallel {\bf k}$), we performed ultrasound experiments for $\pm{\bf B}$ and compared the results.
Up to an ultrasound frequency of 500 MHz, we did not observe reproducible difference with the experimental resolution of $\Delta c/c \sim 10^{-5}$.

\subsection{Field dependence of dielectric constants}

\begin{figure}[tb]
\centering
\includegraphics[width=0.9\linewidth]{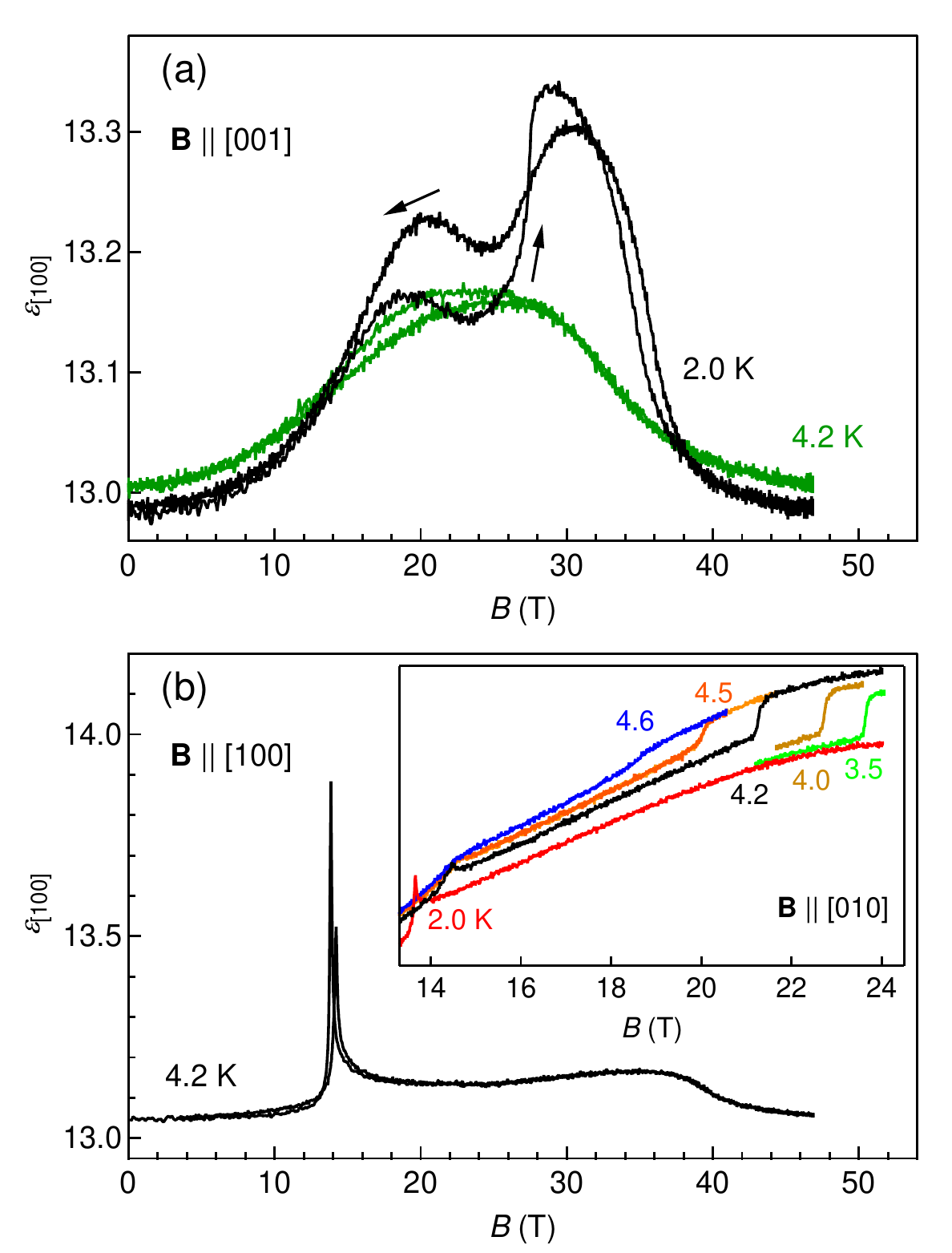}
\caption{
Dielectric constant along the [100] direction $\varepsilon_{[100]}$.
The magnetic fields are applied along the (a) [001] and (b) [100] directions.
The inset shows the results obtained under static fields along the [010] direction at selected temperatures.
} 
\label{dielectric}
\end{figure}

Figures \ref{dielectric}(a) and \ref{dielectric}(b) show the magnetic-field dependences of the dielectric constant for electric fields applied along the [100] axis $\varepsilon_{[100]}$.
For ${\bf B} \parallel [001]$ [Fig.~\ref{dielectric}(a)], $\varepsilon_{[100]}$ shows two broad peak structures at 18 and 28 T.
The hysteresis for field-up and -down sweeps indicates that the temperature changes due to a MCE.
The ME coupling in this system is dominated by the exchange-striction mechanism, where the induced polarization is proportional to $\mathbf{S}_i \cdot \mathbf{S}_j$.
The increase of $\varepsilon_{[100]}$ suggests that the induced polarizations are not canceled between the cupolas in a single magnetic layer.
The second peak appears with a drastic change at $B_{\mathrm{c}1}^{[001]}=28$~T, which is consistent with the ultrasound results [Fig.~\ref{US_pls}(a)].
At $B_{\mathrm{c}2}^{[001]}=29.5$~T, no anomaly was resolved.

For ${\bf B} \parallel [100]$ [Fig.~\ref{dielectric}(b)], $\varepsilon_{[100]}$ shows a sharp peak at $B_{\mathrm{c}1}^{[100]}$ and a broad hump at 39 T which corresponds to the saturation field.
The inset figure shows the results for ${\bf B} \parallel [010]$ at selected temperatures.
In addition to the sharp anomaly at $B_{\mathrm{c}1}^{[100]}=13.8$~T, step-like anomalies are observed at $B_{\mathrm{c}2}^{[100]}=20$--24~T depending on the temperature.
The relatively weak anomaly at  $B_{\mathrm{c}2}^{[100]}$ is consistent with the ultrasound and magnetization results.

\subsection{Magnetocaloric effect}

\begin{figure*}[tb]
\centering
\includegraphics[width=0.98\linewidth]{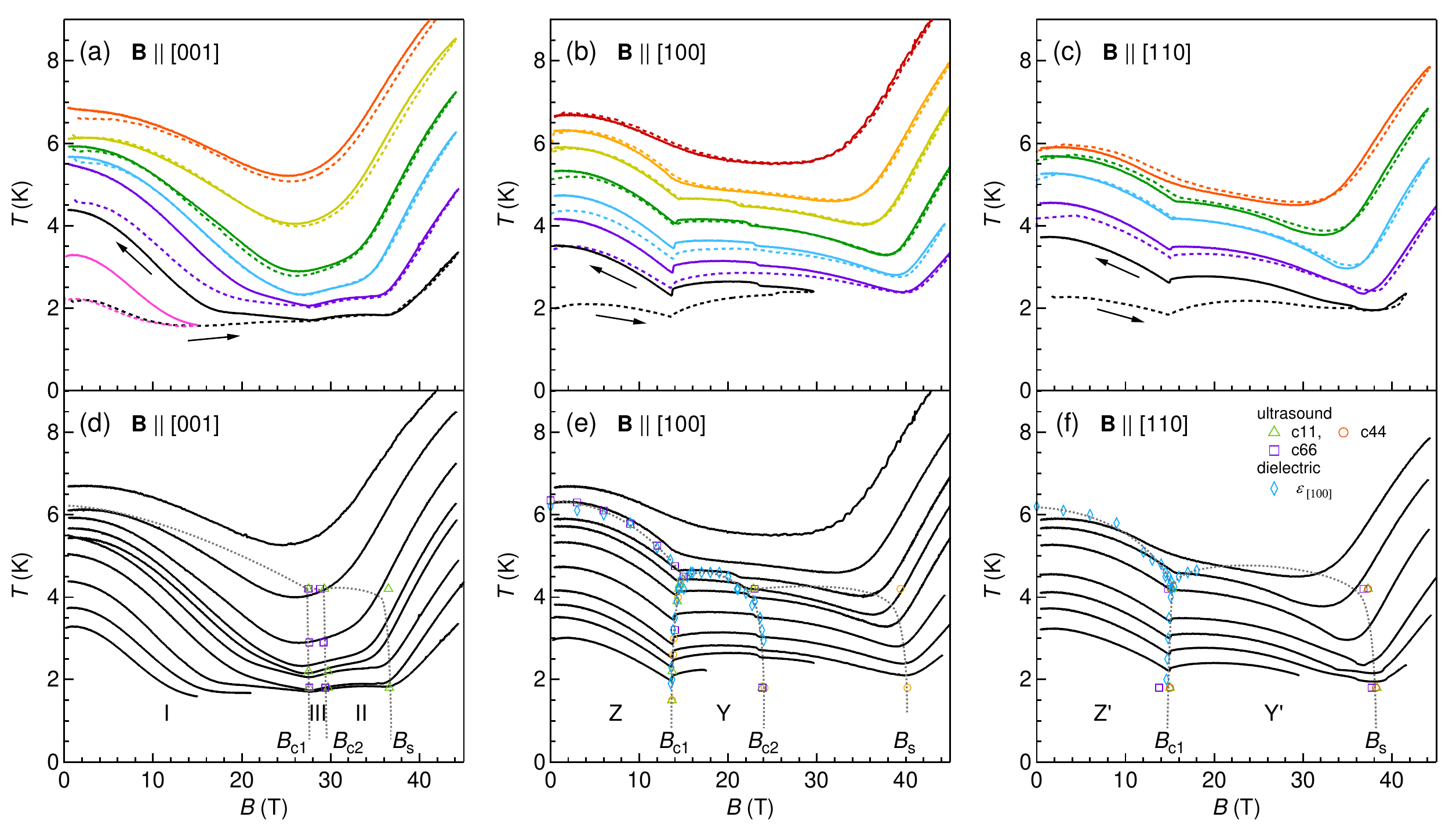}
\caption{
MCE curves obtained for the field directions (a, d) ${\bf B} \parallel [001]$, (b, e) ${\bf B} \parallel [100]$, and (c, f) ${\bf B} \parallel [110]$.
(a-c) Dashed (solid) lines represent the data for field up (down) sweeps.
(d-f) Down-sweep results are shown with the anomalies detected in the elastic and dielectric constants.
Symbols are summarized in Fig.~\ref{MCE}(f).
The dotted lines show guides for the phase boundary and crossover line (see text for details).
} 
\label{MCE}
\end{figure*}

Figures \ref{MCE}(a)--\ref{MCE}(c) show the MCE data for the field directions along [001], [100], and [110] axes, respectively.
The up-sweep (down-sweep) data are shown by dotted (solid) curves.
Here, the temperature change during the adiabatic magnetization is measured from different initial temperatures $T_0$.
For $T_0 >5$~K, the MCE curves are reversible, indicating the adiabatic condition and good thermal contact between the sample and the thermometer.
In contrast, for $T_0 <5$~K, the MCE curves show irreversible behavior with hysteresis.
The irreversible temperature change indicates dissipation such as domain-wall motions or dynamical effects \cite{06PRL_URu2Si2,16JPSJ_Nomura}.
For an overview, only the down-sweep data are presented in Figs. \ref{MCE}(d)--\ref{MCE}(f) with the transition fields determined by the ultrasound and dielectric measurements.

First, the results for ${\bf B} \parallel [001]$ [Figs. \ref{MCE}(a) and \ref{MCE}(d)] are discussed.
When increasing field, the temperature decreases up to $B_{\mathrm{c}1}^{[001]}$.
The temperature decrease indicates an increase of magnetic entropy towards the field-induced transition.
At $B<B_{\mathrm{c}1}^{[001]}$, irreversibilities are is clearly observed.
Above $B_{\mathrm{c}1}^{[001]}$, the MCE curves are reversible and show kinks around the saturation field $B_{\mathrm{s}}^{[001]}$.
No anomaly is observed at $B_{\mathrm{c}2}^{[001]}$, indicating that the entropy does not change at this phase transition. 
Such a transition without entropy change is observed as well in the frustrated magnet CdCr$_2$O$_4$, where also a clear anomaly is detected in the elastic properties \cite{15Zherlitsyn}.
The low-temperature MCE curves merge at $T\approx2$~K at $28<B<36$~T, which suggests a discontinuous change of entropy like a first-order phase boundary in this region.
This might indicate that another phase exists in the low-temperature region, where one needs to perform experiments in static magnetic fields, for instance, with a $^3$He cryostat.

Second, we discuss the results for ${\bf B} \parallel [100]$ [Figs. \ref{MCE}(b) and \ref{MCE}(e)].
For this field direction, three anomalies are observed at $B_{\mathrm{c}1}^{[100]}$, $B_{\mathrm{c}2}^{[100]}$, and $B_{\mathrm{s}}^{[100]}$,  consistent with the ultrasound results.
The MCE curves are irreversible below $B_{\mathrm{s}}^{[100]}$.
Noteworthy, the anomaly at $B_{\mathrm{c}2}^{[100]}$ evidences a first-order phase transition.
At first-order transitions, a reversible temperature change $\Delta T_\mathrm{rev}$ and irreversible heating $\Delta T_\mathrm{irr}>0$ are involved \cite{06PRL_URu2Si2,16JPSJ_Nomura,Tsujii,Sugiura}.
At around $T_0=5$~K, the phase transition at $B_{\mathrm{c}2}^{[100]}$ is detected only for the down sweep.
At the lowest temperature of $T_0=2$~K [black curve in Fig.~\ref{MCE}(b)], in contrast, a temperature increase is observed both for the up and down sweep.
This indicates that $\Delta T_\mathrm{irr}$ is larger than $\Delta T_\mathrm{rev}$ in this temperature range.
Usually, $\Delta T_\mathrm{irr}$ and the hysteresis become smaller at higher temperatures because the energy barrier between two phases can be overcome by thermal fluctuations \cite{06PRL_URu2Si2}.
At around $T_0=5$~K (cyan and green curves), the contributions of $\Delta T_\mathrm{rev}$ and $\Delta T_\mathrm{irr}$ seem to be cancelled out for the up sweep ($\Delta T_\mathrm{irr}>0$ and $\Delta T_\mathrm{rev}<0$), while summed up for the down sweep ($\Delta T_\mathrm{irr}>0$ and $\Delta T_\mathrm{rev}>0$).

Third, we discuss the results for ${\bf B} \parallel [110]$ [Figs. \ref{MCE}(c) and \ref{MCE}(f)].
At least, two anomalies are observed at $B_{\mathrm{c}1}^{[110]}$ and $B_{\mathrm{s}}^{[110]}$.
Hysteresis appears below $B_{\mathrm{s}}^{[110]}$ similar to ${\bf B} \parallel [100]$.
At around 35~T, below the saturation, tiny kinks are observed.
These anomalies are not seen in the ultrasound measurements.
Apparently, these anomalies depend on the peak field (i.e., different $dB/dt$ rate) and are not well reproduced.
They might be related to another phase transition, however, we leave it for a future work.

\section{Theoretical phase diagram}
Our experimental results show two additional features at $B_{\mathrm{c}2}^{[001]}$ and $B_{\mathrm{c}2}^{[100]}$, which are not predicted in previous theoretical calculations \cite{Kato19}.
One possible origin is a chiral twist of the cupola [Fig.~\ref{fig1}(b)], which tilts the DM vectors as well.
In this section, we present a theoretical phase diagram for a twist angle of $\phi=1^{\circ}$ and discuss how the phase diagram changes from the case $\phi=0^{\circ}$ \cite{Kato19}.
For representative spin orientations, see Appendix A.

Figures \ref{PD}(a)--\ref{PD}(c) show the results for $\phi=1^{\circ}$ for ${\bf B} \parallel [001]$, ${\bf B} \parallel [100]$, and ${\bf B} \parallel [110]$.
We find that all the phases seen for $\phi=0^{\circ}$ are also stable for $\phi=1^\circ$, namely, I, II, III, Y, Z, Y', and Z' \cite{Kato17,Kato19}.
These phases are identified by ${\bf m}_{\rm AF}$.
In each phase, ${\bf m}_{\rm AF}$ behaves qualitatively the same as for the case $\phi=0^{\circ}$, except for the phase III 
where the direction of ${\bf m}_{\rm AF}$ deviates slightly from the $[100]$ direction due to the nonzero $\phi$.
The obtained phase diagrams are very similar for the field directions ${\bf B} \parallel [001]$ and ${\bf B} \parallel [110]$,
while the high-field Y phase is largely affected for ${\bf B} \parallel [100]$:
the phase boundary on the high-field side of the Y phase shifts significantly to low fields, leaving a clear crossover (cyan dashed line) at the original phase boundary.

\begin{figure*}[tb]
\centering
\includegraphics[width=0.9\linewidth]{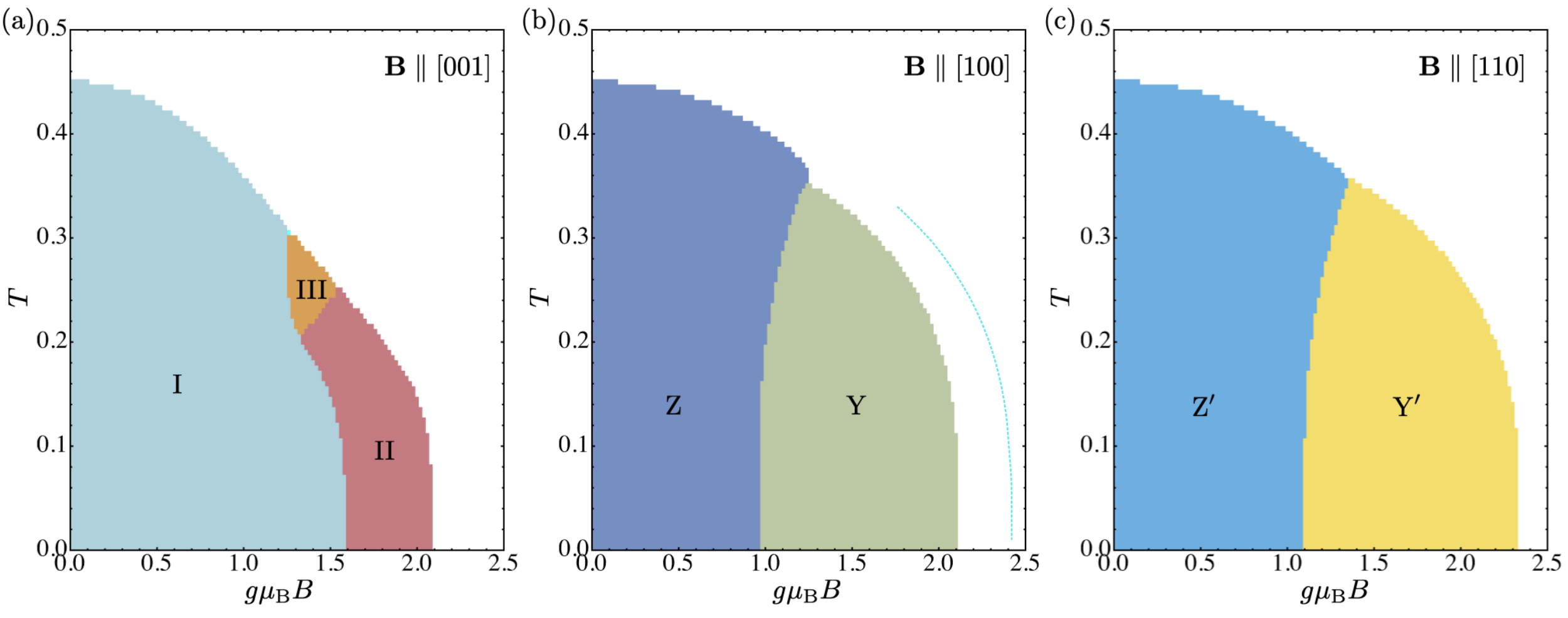}
\caption{
Theoretical magnetic-field-temperature phase diagram for (a) ${\bf B} \parallel [001]$, (b) ${\bf B} \parallel [100]$, and (c) ${\bf B} \parallel [110]$.
Here, the results for the effective Hamiltonian [Eq. %()
\eqref{eq:model}] with the parameters $J_1=0.6$, $J_2=1/6$, $J'=1/2$,  $J''=1/100$, $D=0.7$, $\theta=90^\circ$, and $\phi = 1^\circ$ are shown.
The cyan dashed line in (b) shows the crossover field extracted from an inflection point of $dm/dB$.
For the representative spin structures of the proposed phases, see Appendix A.
} 
\label{PD}
\end{figure*}

\section{Discussions}
Here, we compare the theoretical and experimental results.
Based on the calculated entropy and elastic constants, we discuss the MCE and acoustic results in the field-induced phases.

\subsection{Entropy landscape}

\begin{figure*}[tb]
\centering
\includegraphics[width=0.9\linewidth]{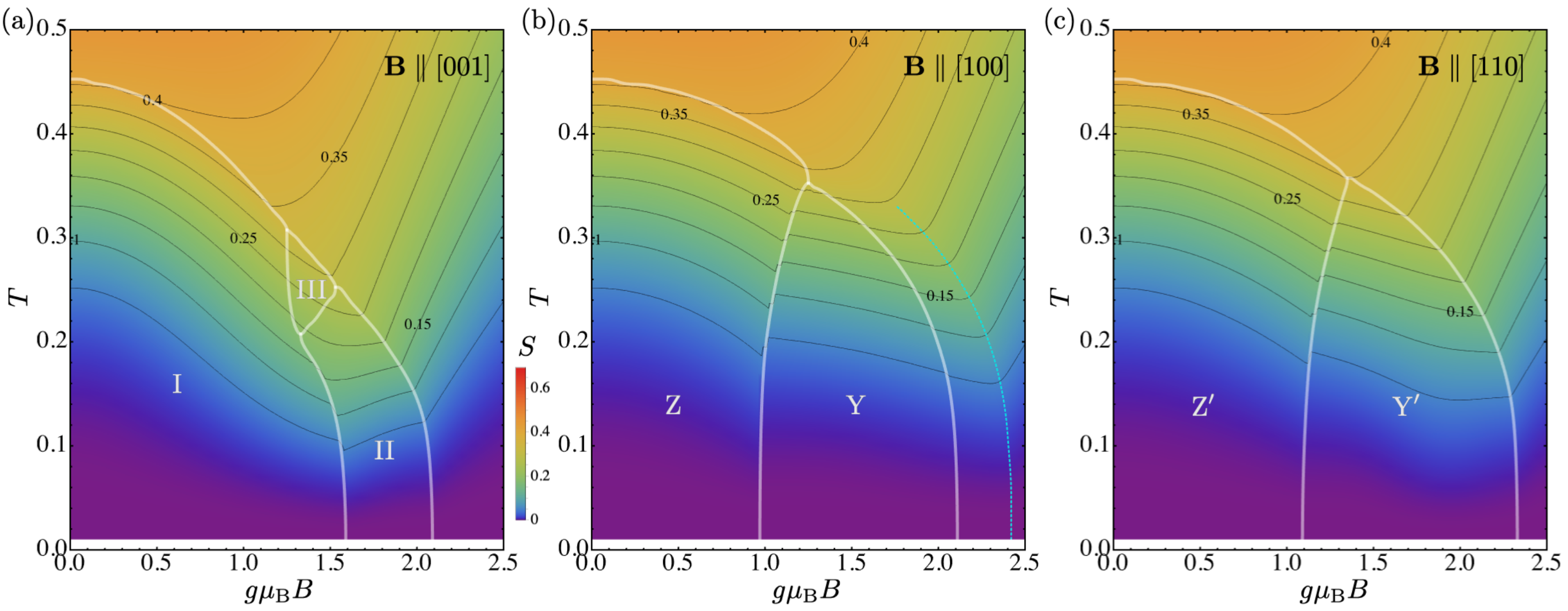}
\caption{
Contour plots of the theoretically calculated magnetic entropy for (a) ${\bf B} \parallel [001]$, (b) ${\bf B} \parallel [100]$, and (c) ${\bf B} \parallel [110]$.
The used parameters are the same as in Fig.~\ref{PD}.
The white solid lines and the cyan dashed line show the phase boundaries and crossover extracted from the phase diagram (Fig.~\ref{PD}), respectively.
} 
\label{entropy}
\end{figure*}

Figures \ref{entropy}(a)--\ref{entropy}(c) show the results of calculations of the entropy landscape in the phase diagram.
The entropy at each field-temperature point is calculated from the temperature dependence of the internal energy as explained in Sec.~\ref{sec:theoretical_procedures}.
Our MCE results [Figs. \ref{MCE}(a)--\ref{MCE}(c)], which follow the isentropic curves [black lines in Figs. \ref{entropy}(a)--\ref{entropy}(c)], agree quite well with the calculations.
For ${\bf B} \parallel [001]$, the entropy decreases toward the phases II and III, and increases above $B_\mathrm{s}$.
The phase III-II boundary is different between the experiment and theory, where the phase III is only stable at finite temperatures.
This discrepancy is discussed in the next subsection.
For ${\bf B} \parallel [100]$, the entropy decreases toward $B_\mathrm{s}$ with slight discontinuous anomalies at $B_\mathrm{c1}$ and $B_\mathrm{c2}$.
For ${\bf B} \parallel [110]$, the entropy decreases toward $B_\mathrm{s}$ with a discontinuous anomaly at $B_\mathrm{c1}$.
The discontinuous entropy changes indicate first-order phase boundaries, which are also consistent with the experimental results (Figs. \ref{US_pls}--\ref{MCE}).

\subsection{Elastic anomalies}

\begin{figure*}[tb]
\centering
\includegraphics[width=0.95\linewidth]{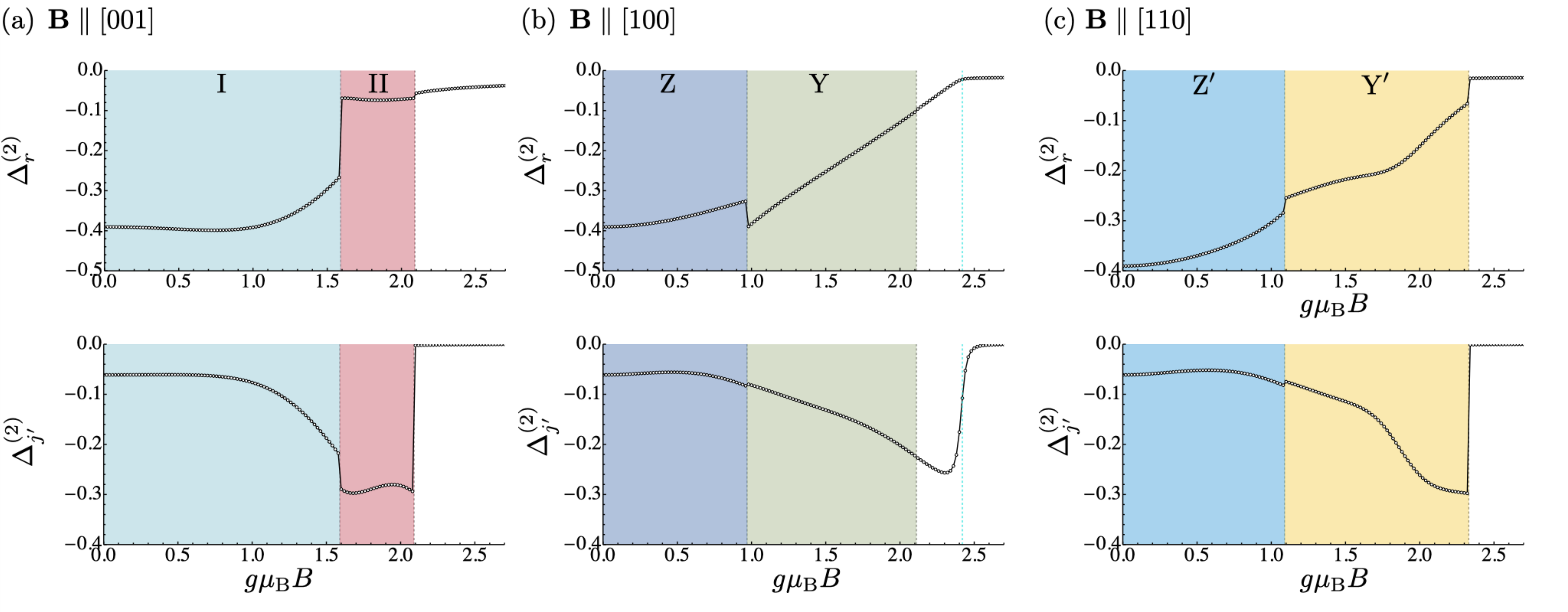}
\caption{
Second derivatives of the magnetic energy with respect to strain as a function of the magnetic field along (a) ${\bf B} \parallel [001]$, (b) ${\bf B} \parallel [100]$, and (c) ${\bf B} \parallel [110]$.
The derivatives are taken at $T=0$ for the strains $\epsilon_{xy}$ (upper panel) and $\epsilon_{xx}$ (lower panel).
For details, see the main text.
} 
\label{derivative1}
\end{figure*}

\begin{figure}[tb]
\centering
\includegraphics[width=0.9\linewidth]{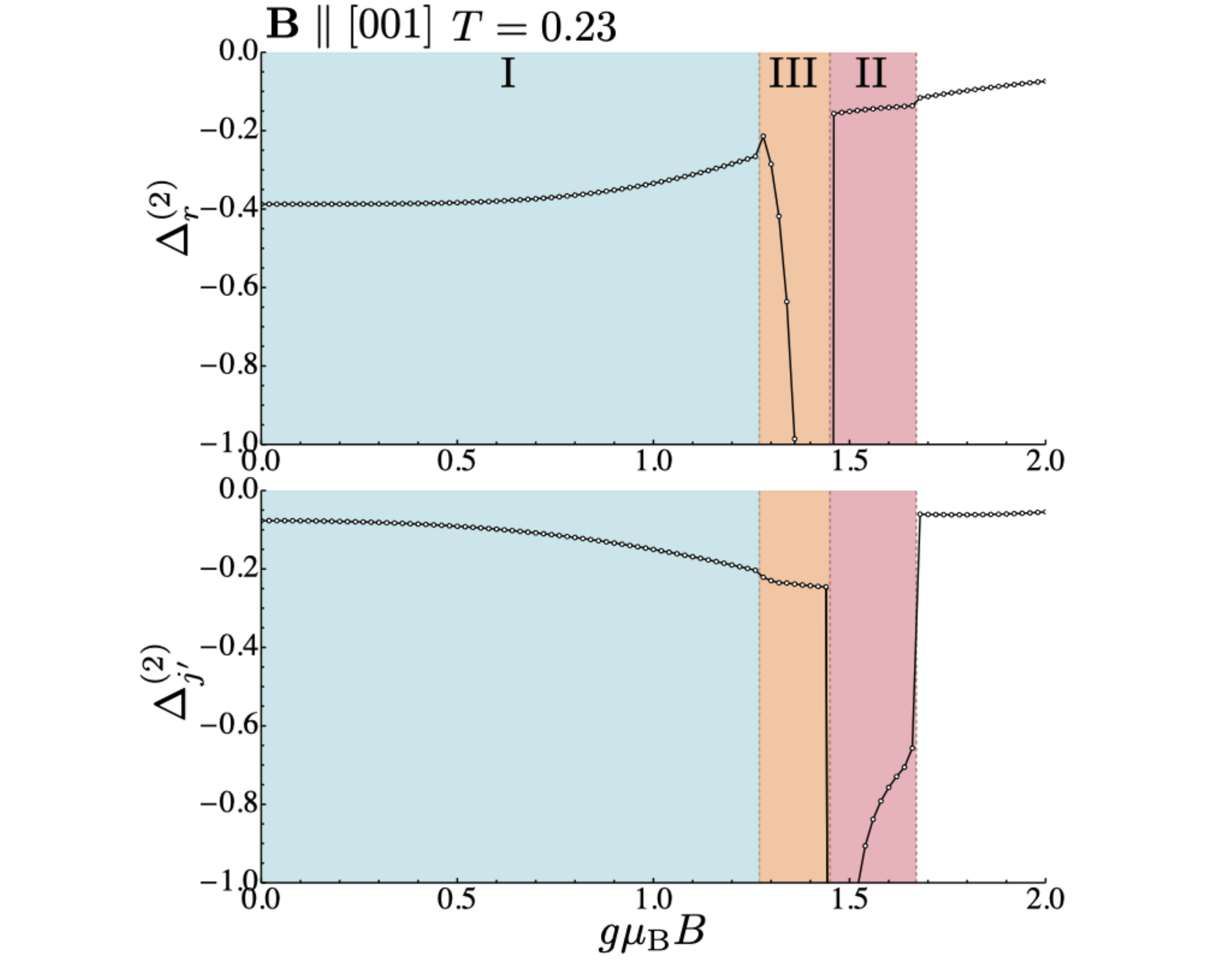}
\caption{
Second derivatives of the magnetic energy near the phase III-II boundary ($T=0.23$) for ${\bf B} \parallel [001]$.
The derivatives are taken for the strains $\epsilon_{xy}$ (upper panel) and $\epsilon_{xx}$ (lower panel).
} 
\label{derivative2}
\end{figure}

Here, we discuss the elastic constants of the field-induced phases based on the CMF theory.
Figures \ref{derivative1}(a)--\ref{derivative1}(c) present the calculated second derivatives of the magnetic free energy with respect to strain, $\Delta_r^{(2)}$ and $\Delta_{j'}^{(2)}$, for three magnetic field directions.
The upper and lower panels show the results for the $\epsilon_{xy}$ ($c_{66}$ mode) and $\epsilon_{xx}$ ($c_{11}$ mode) strains, respectively.
We note that the derivatives for different strains (acoustic modes) clearly show different responses, reflecting the symmetry of the magnetic structures.
Because of this acoustic-mode dependence, ultrasound results provide a powerful basis to discuss the magnetic phases.

Compared with the experimental ultrasound results [Figs. \ref{US_pls}(a)--\ref{US_pls}(c)], the calculated results are in quantitative agreement for ${\bf B} \parallel [100]$ and ${\bf B} \parallel [110]$.
The characteristic curvature changes at $B_{\mathrm{c}1}^{[100]}$, $B_{\mathrm{s}}^{[100]}$, $B_{\mathrm{c}1}^{[110]}$, and $B_{\mathrm{s}}^{[110]}$ are well captured by the calculation.
The small anomaly at $B_{\mathrm{c}2}^{[100]}$ is also reproduced in the calculation.
The relatively broad anomaly at $B_{\mathrm{s}}^{[100]}$, which is predicted to be a crossover, is also reproduced [Figs. \ref{US_pls}(b) and \ref{derivative1}(b)].
These agreements support the magnetic structures predicted by our theoretical model (Z, Y, Z', and Y' phases).

For the case ${\bf B} \parallel [001]$, we need a careful consideration. 
The calculation predicts that the phase III only appears at finite temperatures.
Thus, at the lowest temperature, only two phase transitions are predicted, while our ultrasound experiments detect three phase transitions ($B_{\mathrm{c}1}^{[001]}$, $B_{\mathrm{c}%1
2}^{[001]}$, $B_{\mathrm{s}}^{[001]}$) below 4 K.
Our ultrasound results show a characteristic anomaly at the field range from $B_{\mathrm{c}1}^{[001]}$ to $B_{\mathrm{c}%1
2}^{[001]}$, where only the $c_{66}$ mode exhibits a drastic response, while the $c_{11}$ mode only shows a small anomaly.
Figures \ref{derivative1}(a) and \ref{derivative2} display the derivatives of the free energy across the phases I-II, and I-III-II, respectively.
Only across the phase III, the drastic response in the $c_{66}$ mode is reproduced.
For the case of the phase II, the $c_{11}$ mode shows even larger response near the saturation field.
Therefore, the results suggest that the phase transitions at $B_{\mathrm{c}1}^{[001]}$ and $B_{\mathrm{c}2}^{[001]}$ correspond to the I-III and III-II transitions, respectively.
One discrepancy, the hardening of $c_{66}$ (experiment) and the decrease of $\Delta_r^{(2)}$ (theory) in the phase III (Fig.~\ref{derivative2}) might be due to the crystal deformation in this phase.
The ultrasound results [Fig.~\ref{US_pls}(a)] show the drastic change of the elastic constants in the phase III with large hysteresis, indicating that the crystal structure changes in this phase.
When the magnetic structure changes, the crystal structure can also change due to magneto-elastic coupling, which modifies the exchange coupling to stabilize the magnetic structure at the expense of the elastic energy \cite{Penc04,Miyata20}.
Thus, the exchange parameters used in the CMF calculations might change under magnetic field.
Another partial discrepancy, the theoretical prediction of the phase III stabilized only at finite temperatures, might also be due to the modified exchange parameters under magnetic field.
The stability of the phase III is sensitive to the angle of the DM vector $\theta$ \cite{Kato17}, namely the O$^{2-}$ anion position, which can be modulated via the magneto-elastic coupling.
By including this coupling term in the model Hamiltonian, the agreement between the experiment and theory might further improve.

\section{Conclusions}
We present high-field results on SrTCPO using ultrasound, dielectric, and magnetocaloric-effect measurements.
Compared to previous studies, we find new phase boundaries at $B_{\mathrm{c}2}^{[001]}$ and $B_{\mathrm{c}2}^{[100]}$.
Both phase transitions have not been detected by magnetization measurement.
Based on CMF calculations, we discuss the origin of these phases.
By considering a chiral twist effect of the cupolas, the $B$--$T$ phase diagram is reasonably well reproduced for three magnetic field directions.
The assignments of the field-induced phases are carefully done by comparing the entropy and elastic response obtained by experiments and calculations.
We propose that the phase between $B_{\mathrm{c}1}^{[001]}$--$B_{\mathrm{c}2}^{[001]}$ is the phase III predicted by the CMF theory based on the drastic response of the $c_{66}$ mode.
The reasonable agreement between the experiment and theory indicates that the CMF theory well captures the nature of this quantum spin system, where the competing exchange interactions and the chirality play an important role in the emergence of the rich phase diagram.

%\vspace{3cm}
\bigskip
We acknowledge the support of the HLD at HZDR, member of the European Magnetic Field Laboratory (EMFL), and the DFG through the Collaborative Research Center SFB 1143 (Project No. 247310070) and the W\"urzburg-Dresden Cluster of Excellence ct.qmat (EXC 2147, Project No. 39085490).
This work was partly supported by the JSPS KAKENHI, Grants-In-Aid for Scientific Research (Nos. JP19K23421, JP20K14403, JP22H00104, JP22K03509) and JSPS Bi-lateral Joint Research Projects (JPJSBP120193507)
This work was partly performed at the High Field Laboratory for Superconducting Materials, Institute for Materials Research, Tohoku University (Project No. 18H0014).
T. N. was supported by the JSPS through a Grant-in-Aid for JSPS Fellows.
K. K. was supported by the MEXT Leading Initiative for Excellent Young Researchers (LEADER).
%The crystal structure figures have been created by using the visualization software VESTA

\section*{Appendix A: Spin configurations}
\begin{figure*}[tb]
\centering
\includegraphics[width=0.9\linewidth]{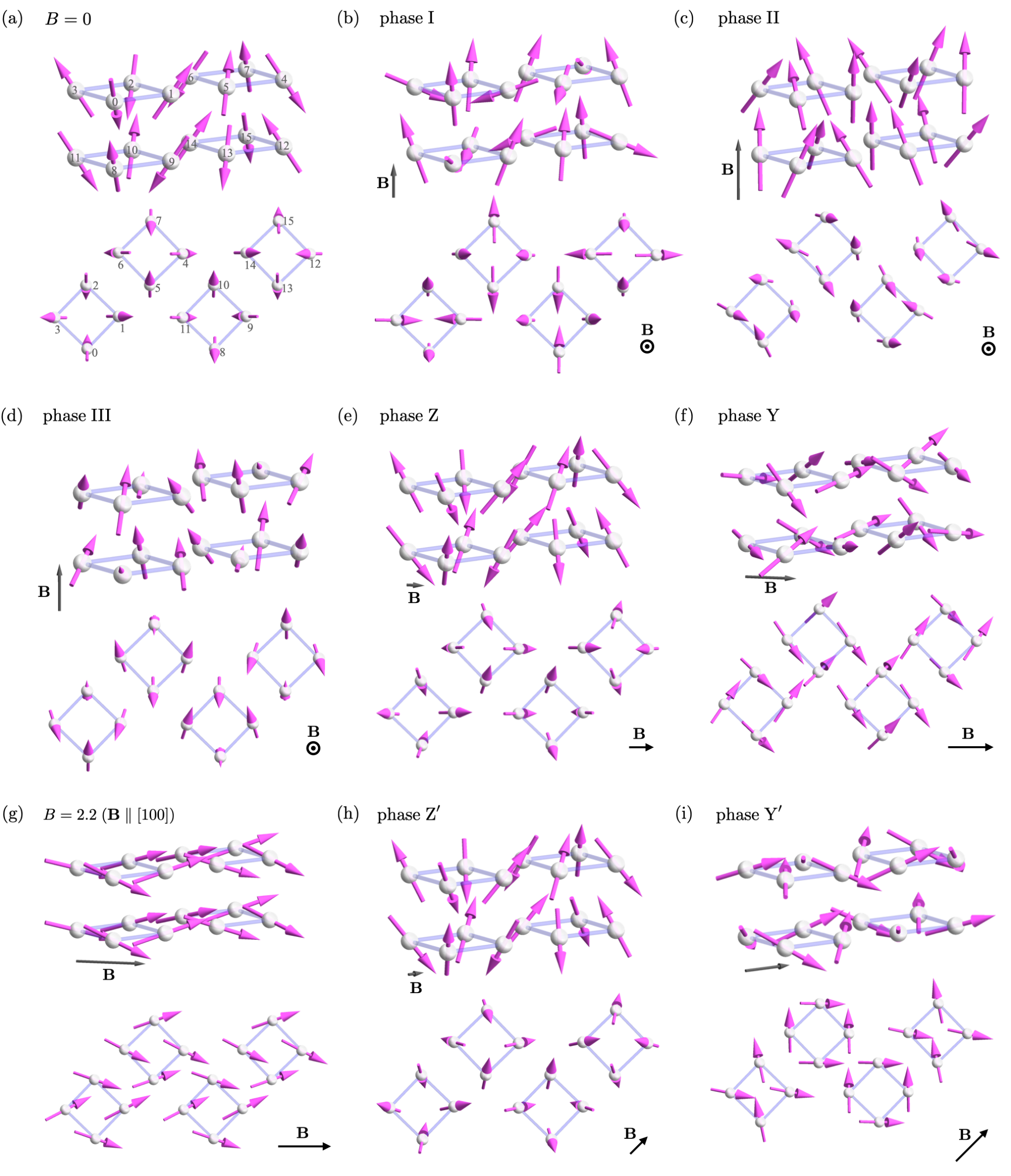}
\caption{
Spin configurations with the following parameters: 
(a) $B=0.0$, $T=0.00$, 
(b) ${\bf B}||[001]$ $B=1.0$, $T=0.00$, 
(c) ${\bf B}||[001]$ $B=1.8$, $T=0.00$, 
(d) ${\bf B}||[001]$ $B=1.3$, $T=0.23$, 
(e) ${\bf B}||[100]$ $B=0.5$, $T=0.00$, 
(f) ${\bf B}||[100]$ $B=1.5$, $T=0.00$, 
(g) ${\bf B}||[100]$ $B=2.2$, $T=0.00$, 
(h) ${\bf B}||[110]$ $B=0.5$, $T=0.00$, and
(i) ${\bf B}||[110]$ $B=1.5$, $T=0.00$.
} 
\label{appendix01}
\end{figure*}

Figure \ref{appendix01} presents the representative spin configurations obtained by the CMF theory including the chiral twist effect. The field and temperature parameters are given in the caption.

\end{document}